# Domain switching and exchange bias control by electric field in multiferroic conical magnet Mn$_2$GeO$_4$

J. K. H. Fischer[1,*], H. Ueda[2], and T. Kimura[1]

[1]*Department of Advanced Materials Science, University of Tokyo, 5-1-5 Kashiwanoha, Kashiwa, Chiba 277-8561, Japan*
[2]*Swiss Light Source, Paul Scherrer Institute, 5232 Villigen-PSI, Switzerland*

The electric field effect on magnetism was examined in the multiferroic conical magnet Mn$_2$GeO$_4$, which shows a strong coupling between ferromagnetic and ferroelectric order parameters. The systematic evaluation of the electric polarization in the multiferroic phase below 5.5 K under various field cooling conditions reveals that small magnetic fields of 0.1 T significantly reduce the required electric fields needed to reach saturation. By applying electric fields during magnetic field dependent hysteresis measurements of magnetization $M$ and polarization $P$ an electrically controllable exchange bias was observed, a phenomenon exceedingly rare in single phase multiferroics. Furthermore, non-reversible electric switching of $P$ and $M$ domains was achieved under specific magnetic field conditions.

## I. INTRODUCTION

Although multiferroics were already discovered 60 years ago in the Soviet Union by Smolenskii *et al.* [1,2], the field has started progressing rapidly after the turn of the millennium [3–5]. Especially the discovery of the giant magnetoelectric effect in magnetically induced multiferroics [6] encouraged more research in this field, since it might enable, among others, fast and energy-efficient magnetic data storage written by an electric field [7]. In order to find applications on a large scale, however, multiferroics are still lacking in some essential properties, like a strong coupling between their magnetic and ferroelectric orders [8].

A plausible device functionality employing the magnetoelectric effect is electric field tuning of an exchange bias. The exchange bias is a phenomenon associated with the exchange anisotropy created at the interface between ferromagnetic (FM) and antiferromagnetic (AFM) layers. It results in a shift of the coercive field in the magnetic hysteresis loop [9,10] and has long been used in spin-valve devices. However, if such an exchange bias effect can be controlled by an electric field, it could find more extensive applications in spintronic devices. As such, it was demonstrated in a multilayer of an AFM magnetoelectric (Cr$_2$O$_3$) and a ferromagnet (e.g., Co/Pt) fifteen years ago [11]. Later on, continuing efforts have been carried out to improve the performance [12,13].

Without using the FM/AFM interface effect, however, the electric field control of exchange bias is achievable by using single phase multiferroics exhibiting strong coupling between ferromagnetic and ferroelectric orders. In fact, an electric-field induced exchange bias in the magnetization hysteresis has been observed very recently in multiferroic hexaferrite systems (Ba$_{0.4}$Sr$_{1.6}$Mg$_2$Fe$_{12}$O$_{22}$ [14] and Ba$_{2-y}$Sr$_y$Co$_2$Fe$_{12-x}$Al$_x$O$_{22}$ [15]) whose multiferroicity is ascribed to spiral spin order. The reverse effect, i.e. an exchange bias induced in the ferroelectric polarization hysteresis by applying a magnetic field, was observed in multiferroic orthoferrite Dy$_{0.7}$Tb$_{0.3}$FeO$_3$ [16].

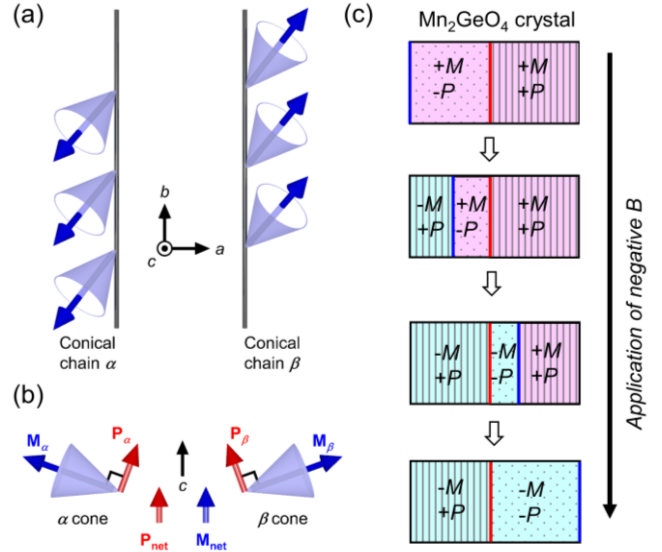

FIG. 1. Simplified schematic of the conical magnetic structure in Mn$_2$GeO$_4$. (a) The view along the $c$ axis and (b) the side view. The structure consists of two conical spin chains along the $b$ axis ($\alpha$ and $\beta$). Local magnetization $\mathbf{M}_{\alpha,\beta}$ (blue arrows) is aligned with the cone axis, while the magnetically induced polarization $\mathbf{P}_{\alpha,\beta}$ is perpendicular. The resulting magnetization $\mathbf{M}_{net}$ and polarization $\mathbf{P}_{net}$ both point along the $c$ axis due to a canted antiferromagnetic coupling between the two chains. (c) Schematic illustrations of magnetoelectric inversion of the ferroelectric domain pattern observed in Ref. [17]. Pink and light blue areas denote $+M$ and $-M$ domains, respectively. Striped and dotted areas denote $+P$ and $-P$ domains, respectively. Red lines represent a ferroelectric domain boundary formed during cooling, whose position is unchanged by applying a magnetic field $B$. Blue lines denote a ferromagnetic domain boundary which develops by the application of negative $B$ and clamps with the ferroelectric one.

Here we examine the electric field effect on magnetization hysteresis loops, i.e. electrically controllable exchange bias in a unique magnetically induced multiferroic Mn$_2$GeO$_4$. This compound crystallizes in the *Pnma* orthorhombic structure and exhibits both ferromagnetism and ferroelectricity below 5.5 K [18]. The multiferroic nature is ascribed to conical spiral magnetic order. The magnetically induced polarization $P$ in transverse conical magnetic systems usually develops in the direction normal to the magnetization $M$ [19] through the inverse Dzyaloshinskii-Moriya (DM) interaction [20]. However, $P$ in $Mn_2GeO_4$ [18] is *parallel* to $M$, a unique feature among the spiral spin-ordered multiferroics [21]. This is resulting from two transverse conical chains which exist in its multiferroic phase, as illustrated in Fig. 1(a) [22]. The Mn moments in each chain make up the cones, whose angles with the $a$, $b$, and $c$ axes are finite. The conical chains $\alpha$ and $\beta$ are canted-antiferromagnetically coupled. For the magnetization, the moments along the $a$ and $b$ axes are cancelled out, while a finite moment along the $c$ axis remains. Through the inverse DM effect [20] the cycloidal components in each chain result in a local $P$ in the direction perpendicular to the propagation vector $\|b$, and the spin rotation axis. The vector sum of these local $P$ in the $\alpha$ and $\beta$ chains ($\mathbf{P}_\alpha$ and $\mathbf{P}_\beta$) results in $\mathbf{P}_{net}$ along the $c$ axis. Thus, both a net magnetization $\mathbf{M}_{net}$ and a net electric polarization $\mathbf{P}_{net}$ develop along the $c$ axis, as shown in Fig. 1(b). In $Mn_2GeO_4$, it has been shown that ferroelectric and ferromagnetic domain walls are strongly clamped via the DM interaction [18]. Ferroelectric and ferromagnetic switching occur synchronized, with corresponding coercive fields, by a flop of the cone axis [22]. Recently, a striking magnetoelectric inversion of the ferroelectric domain pattern has been observed by optical second harmonic generation (SHG) [17]. Schematic illustrations of the domain inversion reported in Ref. [17] are displayed in Fig. 1(c). Here, the *pattern* of the ferroelectric domains stays invariant while the *sign* of the domains, i.e. the direction of $P$ in each domain, is reversed by an external magnetic field. This originates from the interactions of the AFM order parameter, magnetization $M$, and polarization $P$ involved in the multiferroic phase of $MnGeO_4$. Cooling under an external magnetic field $B$ will lead to $M$ being uniform while a distribution of $P$ is possible. However, the AFM order parameter is invariant under application of $B$ so that the free energy can only be minimized if $P$ switches sign simultaneously with $M$. For details of the phenomenological Landau theory see Ref. [23].

In this paper, we report on an electrically controllable exchange bias in magnetization hysteresis loops of $Mn_2GeO_4$. We demonstrate electric switching of polarization and magnetization, which had so far been elusive in $Mn_2GeO_4$. To achieve this, it is necessary to "bias" both polarization and magnetization loops with a magnetic field to the point in the hystereses where the respective property is about to change sign. There a changing of the electric field $E$ from negative to positive (or vice versa) leads to a (non-reversible) switching of the sign of $P$ and $M$.

## II. EXPERIMENTAL DETAILS

Single crystals of $Mn_2GeO_4$ were synthesized by the floating zone method as reported in Ref. [18]. The crystals were oriented by Laue diffraction and cut into thin plates with approximate sizes of 4 mm² × 200 µm and the widest faces normal to the $c$ axis. Silver electrodes were applied on the widest faces of the samples. In this study, both electric field $E$ and magnetic field $B$ were applied along the ferroelectric/ferromagnetic $c$ axis.

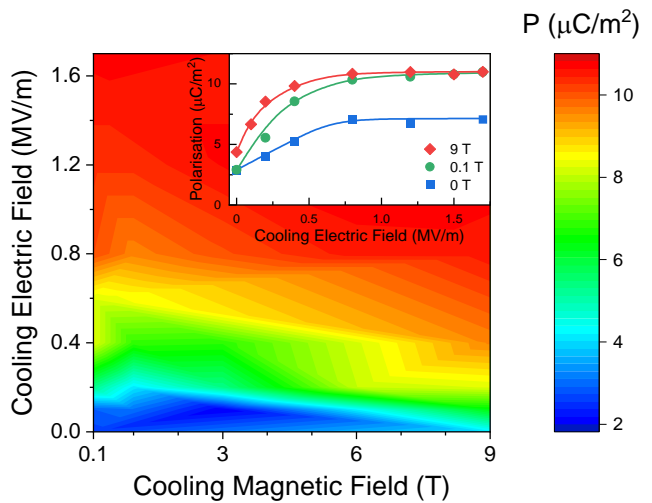

FIG. 2. Color map of the dependence of the spontaneous polarization on cooling electric and magnetic fields. The lowest values are shown in dark blue, while the highest values are deep red. Inset: Comparison of saturation polarization in dependence of applied electric fields for three applied magnetic fields, no magnetic field (blue squares), a small magnetic field of 0.1 T (green circles) and the largest value of 9 T (red diamonds).

The electric polarization $P$ was obtained by integrating the displacement current measured during sweeping temperature or $B$ with an electrometer (Keithley 6517). The magnetization $M$ was measured with the AC susceptibility and DC magnetization (ACMS) option of a Physical Property Measurement System (PPMS, Quantum Design). To control temperature and apply magnetic fields, the PPMS was employed. To examine the electric field effect on $M$ and $P$, a home-made insert, which allows the application of $E$ during measurements of $P$ and $M$, was installed in the PPMS. On the insert, the silver electrodes on a sample were electrically connected with a voltage source (Keithley 6517).

## III. RESULTS AND DISCUSSION

### A. Systematic evaluation of cooling fields

In $Mn_2GeO_4$, very large electric fields are necessary to reach the saturation value of the polarization. Over 2 MV/m are required [22], compared to less than 0.2 MV/m in other spiral spin ordered multiferroics, e.g., in the cycloidal system $Ni_3V_2O_8$ [24]. Here we show, by a systematic evaluation of cooling electric and magnetic fields, that the application of small magnetic fields suffices to reduce the required electric field significantly.

We performed a systematic evaluation of cooling conditions in order to find the highest saturation polarization in $Mn_2GeO_4$. For the systematic evaluation, $P$ of four samples was obtained with measurements of pyroelectric (upon heating) and magnetoelectric (between ±1 T at 4.5 K) current in over 100 combinations of cooling electric and magnetic fields. Each measurement involved cooling from high temperature (60 K) into the multiferroic phase below 5.5 K, with the intended electric and magnetic fields only applied during cooling.

Figure 2 displays the contour plot of $P$ versus the strength of $E$ and $B$ upon cooling. The $P$ values were taken after removing $E$ and $B$, and therefore correspond to those of spontaneous polarization in the respective cooling conditions. In the color map the highest $P$ values are shown in deep red, while the smallest are colored dark blue. The result shows that the required electric field to fully pole the samples can be reduced from over 2 MV/m [22] to 0.8 MV/m by applying a small magnetic field of 0.1 T. The inset of Fig. 2 accentuates this point by comparing the $P$ values under no magnetic field (blue squares), 0.1 T (green circles), and 9 T (red diamonds). While the zero magnetic field curve reaches a significantly lower $P$ value, the measurements taken with a cooling magnetic field of 0.1 T and low electric fields reach only slightly lower values than those taken with the maximum cooling field of 9 T. Only when cooling with both $E$ and $B$ will the $P$ and $M$ domains be fully coupled so that $P$ can be driven into a single domain state by $B$. When cooling without electric field a multidomain state of $P$ with a distribution of domains is retained, so that the saturation value of $P$ is also randomly distributed in each measurement. As can be clearly seen in the inset of Fig. 2 only a value of <5 μC/m² can be reached in an $E = 0$ cooled sample, compared to the maximum value of $P \approx 11$ μC/m². Even applying large magnetic and electric fields in the multiferroic phase below 5.5 K cannot increase this value since $M$ and $P$ were not coupled upon cooling, so that $+P$ and $-P$ states are possible while $M$ is in a single domain state. This can be clearly appreciated by viewing the ferroelectric domain pattern observed by optical SHG in Ref. [17] or its schematic depiction in Fig. 1(c). We also recreated the conditions of Ref. [17], $B_{cool} = -230$ mT and $E_{cool} = 0$. As expected, the highest saturation values of $P$ cannot be reached with these cooling fields. Evidently, $P$ remains in the multidomain state visualized by SHG. Conversely, cooling with $B = 0$ will again lead to decreased values of $P$, since even the highest applied electric fields are not enough to reach saturation, i.e. a single domain state of $P$. Also note that upon cooling with zero magnetic field, the $P$, $M$, and AFM domains are not yet in equilibrium after entering the multiferroic phase. To reach a stable relation a first complete reversal of the magnetization from $+M$ to $-M$ is required [25].

For the subsequent measurements of electric biasing and field switching, the samples were cooled with appropriately high electric and magnetic fields to enter a single domain state of $M$ and $P$.

### B. Electrically controllable exchange bias

Since the $P$ and $M$ domains in $Mn_2GeO_4$ are strongly coupled [17,23], we expect electric fields to be able to affect not only the polarization but also the magnetization hystereses. This is because the $+M$ domain state coupled with the $+P$ domain state will be stabilized by the application of $+E$. After field-cooling from 60 K with adequately high magnetic and electric fields, 9 T and 1 MV/m respectively, we check for effects of applied electric fields during magnetic field sweeping. In the same way that $P$ can be switched by $B$, the invariance of the AFM order parameter leads to the free energy being minimized only if $M$ switches sign simultaneously with $P$ under an applied electric field. In the case of $Mn_2GeO_4$ a $P$-$E$ hysteresis could not be measured since sufficiently high electric fields cannot be applied. However, applying large positive or negative electric fields during sweeping of the magnetic field, lead to the shifts of the hystereses discussed in the following.

Figure 3 shows the effects of electric field biasing with ±1 MV/m for (a) the $P$-$B$ and (b) the $M$-$B$ hysteresis loops.

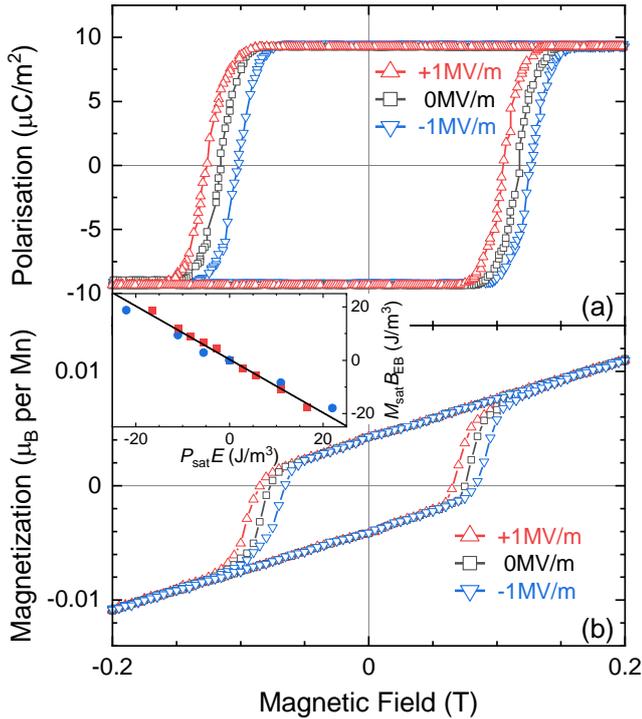

FIG. 3. Electrically controlled exchange bias of (a) polarization and (b) magnetization after field-cooling from 60 K to 4.5 K with $B = 9$ T and $E = 1$ MV/m. Positive electric fields (red upward triangles) shift the $M$ and $P$ hystereses to the left by about 0.01 T compared to no applied field (black squares), while negative electric fields (blue downward triangles) lead to a shift of the same magnitude in the opposite direction. (Inset) Evaluation of bias of magnetic field at different electric fields for magnetization (blue circles) and polarization (red squares). The application of $E$ modulates the ground state energy of $M$ due to the electrostatic energy $P_{sat}E$. Exchange bias shift is therefore linearly dependent on $E$: $B_{EB} = -P_{sat}E/M_{sat}$ (black line).

Compared to the unbiased (0 MV/m, black squares) curves, the curves under applied positive electric fields (1 MV/m, red upward triangles) are shifted to negative values on the magnetic field axis in both the *P-B* and the *M-B* curves, see Fig. 3(a) and 3(b), respectively. The opposite is true for negative applied electric fields (-1 MV/m, blue downward triangles), which shift the curves towards positive magnetic field values. These data clearly show that the exchange bias effect is induced by the application of $E$ in $Mn_2GeO_4$.

Note that there is a slight difference of the coercive fields of polarization and magnetization despite the fact that the same sample was used in both the *P-B* and *M-B* measurements. We presume this deviation to be caused by the different changing rates of $B$ in the two measurements. This assumption is confirmed by *P-B* measurements with decreased $B$ sweeping rates, which reduce the coercive field (not shown).

The inset of Fig. 3 shows a summary of several electric field biasing experiments and leads to the explanation of the electric field induced exchange bias in $Mn_2GeO_4$. As can be expected, larger electric fields $E$ lead to larger shifts on the magnetic field axis, with $B_{EB}$ being the difference between the $E = 0$ and applied electric field curves. For the inset these values were multiplied with the maximum saturation polarization $P_{sat}$ and magnetization $M_{sat}$ values (at 0 T), respectively, in order to compare the involved energies. In accordance with the interpretation of the electrically-induced exchange bias phenomenon given in Ref. [14] the shift along the magnetic field axis $B_{EB}$ results from a modification of the ground state free energy due to the applied $E$. An applied positive $E$ will favor $+P$ and therefore also $+M$ states, which results in a larger coercive magnetic field being required in order to compensate $E$. Due to the clamped $P$ and $M$ domains, both get equally affected by applied electric fields and switch in unison. Therefore, a positive (negative) electric field shifts the polarization and magnetization hystereses to negative (positive) magnetic fields, i.e. to the left (right). The electrostatic energy $P_{sat}E$ is completely compensated by $B_{EB}$ so that: $-M_{sat}B_{EB} = P_{sat}E$. Thus, the shift of the *M-B* hysteresis is linearly dependent on the applied $E$, which is demonstrated by the linear behavior seen in the inset of Fig. 3. Obviously, the shift of the *P-B* hysteresis is also linearly dependent on the applied $E$. In order to compare the $B_{EB}$ values in the *M-B* and *P-B* hystereses we multiplied both with the same $M_{sat}$ value, which is arbitrary in the case of *P-B* but of course does not affect the interpretation.

The observation of an electric-field induced exchange bias clearly demonstrates the strong coupling between $M$ and $P$ in $Mn_2GeO_4$. Compared to the two recent reports showing electrically controlled exchange bias effects in hexaferrites [14,15], the absolute value of $B_{EB}$, and therefore the coercive field, in $Mn_2GeO_4$ is remarkably larger, 0.02 T at 2 MV/m, compared with the previous highest value of <0.01 T at 5 MV/m in $BaSrCo_2Fe_{11.1}Al_{0.9}O_{22}$ [15].

As mentioned above, similar exchange bias behavior has, to our knowledge, so far been observed in only a few systems[14–16]. This stems mainly from the rarity of type-II multiferroics that exhibit ferromagnetism and ferroelectricity.

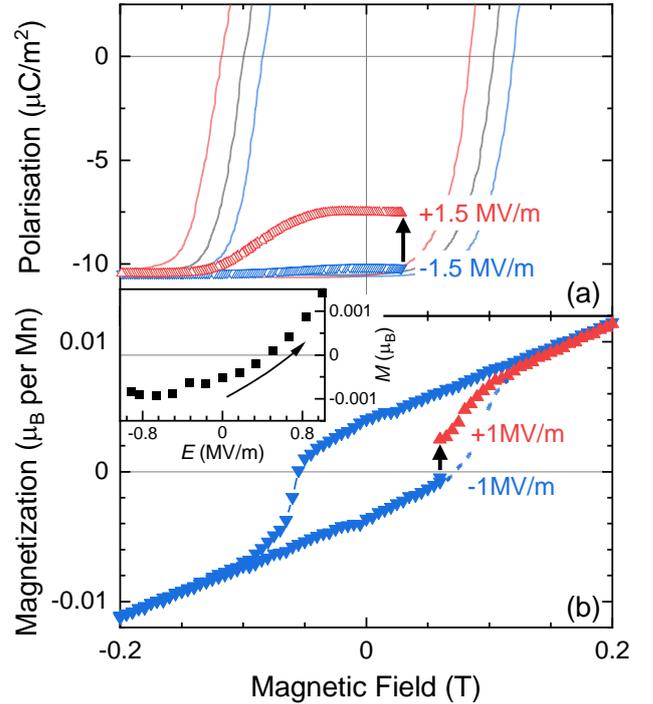

FIG. 4. (a) Electric switching of the polarization at 0.03 T. The data were taken at 4.5 K. The polarization domains can be partially switched by changing the applied electric field from negative to positive. The red triangles show the change in $P(B)$ when decreasing the magnetic field after *E*-switching, equating to 15% (3 μC/m$^2$) of the domains. The blue downward triangles show that the polarization remains nearly unchanged when the electric field is not switched. (b) Electric switching of the magnetization at 4.5 K. After applying negative magnetic and electric fields the magnetization is non-reversibly switched at 0.06 T by changing the electric field from -1 MV/m (blue downward triangles) to +1 MV/m (red upward triangles). (Inset) Electric field dependence of the magnetization: when the applied electric field is increased to above 0.5 MV/m the magnetization crosses zero.

Moreover, the polarization of magnetically induced multiferroics is often comparatively small, exacerbating the observation electrically induced exchange bias. We believe this intriguing behavior deserves further attention, especially considering possible applications of electrically controllable exchange bias and increased coercive fields in spintronic devices.

## C. Electric switching of polarization and magnetization

Comparing the curves where negative (blue) and positive (red) electric fields are applied in Fig. 3, one has to conclude that switching the polarity of $E$ should also induce the reorientation of polarization and magnetization domains. Especially below 0.1 T on the cusp of the rapid increases in $P$ and $M$, where the difference between negative and positive applied electric fields is large. This encouraged us to reattempt electric-field switching of polarization and magnetization, which had so far not been achieved in Mn$_2$GeO$_4$.

In fact, Fig. 4(a) reveals that a partial switching of the polarization by a change of the electric field from -1.5 MV/m to +1.5 MV/m is possible. To accomplish this the magnetic field is swept from +1 T to -1 T and back to +0.03 T while an electric field of -1.5 MV/m is applied. Subsequently, the applied electric field is abruptly changed to +1.5 MV/m (thick black arrow), after which the magnetic field is decreased again to -1 T (red upward triangles). This leads to a significant displacement current, corresponding to a polarization of about 3 μC/m$^2$ (see the length of the thick black arrow). Thus, a change of the applied electric field from large negative to large positive values at +0.03 T prompts a reorientation of about 15% of polarization domains. To make sure that this domain switching is truly caused by the electric field, we first observed what happens if $E$ is kept at its initial value of -1.5 MV/m. In this case sweeping of the magnetic field from +0.03 T back to -1 T leads to only a marginal part of the polarization being switched (blue downwards triangles), i.e. the peak in the displacement current is almost indistinguishable from the background noise. The same is true after waiting for several minutes and when applying no electric field (not shown).

Similarly, Fig. 4(b) shows the switching of the magnetization by changing the applied electric field. In order to achieve this jump from negative to positive magnetization values, it is necessary to go through the $M$-$B$ loop from a large positive magnetic field (+0.5 T) via negative magnetic field (-0.5 T) to the point where the value of the magnetization is only slightly below zero (+0.06 T), i.e. go through about ¾ of the $M$-$B$ hysteresis loop. At +0.06 T, successively increasing the applied electric field from -1.0 MV/m to +1.0 MV/m leads to a crossover from negative to positive magnetization values (see inset), i.e., electric switching of the magnetization. This switching behavior can, however, not be reversed by decreasing the electric field again: once switched the magnetization stays positive and only slightly decreases when negative electric fields are reapplied. When comparing to the exchange bias measurement shown in Fig. 3, it can be seen that the magnetization curves here follow the same behaviors depending on the applied electric fields. The difference here being that the electric field is changed during a pause in the magnetic field sweeping, which allows the electric field to be responsible for the final magnetization switching "impulse". We propose that the comparative weakness of the electric field, i.e. its small effect even compared to tiny magnetic fields, results in the switching being non-reversible and only possible under the right magnetic field conditions. However, the switching behavior can also be observed at other temperatures in the multiferroic phase and on the opposite sides of the hystereses (near 0.05 T), when changing the applied electric field from positive to negative (not shown).

As one can expect after examining the difference in each of the $P$-$B$ and $M$-$B$ curves under varying applied electric fields in Fig. 3, a partial switching of $P$ and $M$ by electric fields is possible. Albeit, for this process "magnetic biasing" is necessary, i.e. magnetic sweeping of about ¾ of the hystereses loops to the $B$ value where $P$-$B$ and $M$-$B$ are just about to drastically increase. Thus, while full $P$-$E$ and $M$-$E$ curves remain unachievable due to the relative weakness of the applied electric fields, it is possible to partially switch magnetization (and polarization) domains purely electrically.

## V. CONCLUSIONS

The electric field effect on magnetism was investigated in the multiferroic conical magnet, Mn$_2$GeO$_4$, showing a strong coupling between ferromagnetic and ferroelectric order parameters. Our experiments have revealed an electrically controllable exchange bias in both the $P$-$B$ and $M$-$B$ hystereses, a phenomenon usually observed in the AFM/FM interface and exceedingly rare in single-phase compounds. The presence of this effect once again demonstrates the striking clamping of polarization and magnetization domains, which is necessary for the applied electric fields to influence the ground state free energy of $M$. The absolute electric-field-induced change of the coercive magnetic field is the highest observed so far, about 0.02 T at $E = 2$ MV/m. Remarkably, we were also able to demonstrate an electric switching of polarization and magnetization, which has so far been elusive in Mn$_2$GeO$_4$. To facilitate this it is necessary to first perform magnetic sweeping of the hystereses loops to where the $P$-$B$ and $M$-$B$ curves are most susceptible to a change in applied electric fields from large negative to large positive values (or vice versa).

The magnetoelectric coupling moderated by an invariant antiferromagnetic order parameter in Mn$_2$GeO$_4$ invites further investigations to understand the so-far only phenomenologically understood mechanism. Since the phenomenon of electrically controllable exchange bias is likely present in any magnetoelectric materials with strong coupling of $P$ and $M$, further insight into what leads to this strong coupling is required. Both exchange bias and control of coercive fields are valuable properties for potential application, hence this phenomenon certainly deserves further attention in future studies.

## ACKNOWLEDGMENTS

The authors thank K. Kimura and R. Misawa for assistance with the experimental setup. This work was supported by JSPS KAKENHI (Grand Numbers JP18F18803, JP17H01143, and JP19H05823).

_______________

*Corresponding author: jfischer@k.u-tokyo.ac.jp